\documentclass{ws-procs961x669}            
\begin{document}
\title{Searching for pulsars in globular clusters with the MeerKAT Radio Telescope}

\author{F. Abbate on behalf of the MeerTIME/TRAPUM Collaboration}

\address{Max Planck Institut f\"ur Radioastronomie,\\
Auf dem H\"ugel 69, D-53121, Bonn, Germany\\
$^*$E-mail: abbate@mpifr-bonn.mpg.de}

\begin{abstract}
Globular clusters are known to host an unusually large population of millisecond pulsar when compared to the Galactic disk. This is thanks to the high rate of dynamical encounters occurring in the clusters that can create the conditions to efficiently recycle neutron stars into millisecond pulsars. The result is a rich population of pulsars with properties and companions difficult or impossible to replicate in the Galactic disk.
For these reasons, globular clusters have been and still are a prime target of searches for new and exciting pulsars. Because of their large distances, the limiting factor inhibiting these discoveries is the telescope sensitivity. The MeerKAT radio telescope, a 64-dish interferometer in South Africa, guarantees unrivalled sensitivity for globular clusters in the southern sky.
Observations of well-studied globular clusters with MeerKAT have already returned more than 35 new pulsars with many more expected. These exciting discoveries will help us to understand more about the neutron star equation of state, stellar evolution, accretion physics and to hunt for intermediate mass black holes. In this talk I will present the prospects and current discoveries of the globular cluster working group in the MeerTIME and TRAPUM programmes.
\end{abstract}


\bodymatter

\section{Pulsars in globular clusters}\label{sec:pulsars}

Globular clusters contain a very rich population of pulsars with 232 known in 36 cluster\footnote{An up-to-date count can be found in the webpage \url{https://www3.mpifr-bonn.mpg.de/staff/pfreire/GCpsr.html}} and this population  is quite peculiar when compared with the one in the Galactic disk. The abundance of pulsars per unit of stellar mass in globular clusters is $\sim 100$ times higher than that of the disk and the vast majority of the pulsars are  millisecond pulsars (MSP) with rotation periods shorter than a few tens of ms. 

The difference between the population of pulsars in globular clusters and in the Galactic disk is caused by the starkly different environments. Globular clusters have no sign of recent star formation and have very high stellar densities in their cores that can reach up to $\sim 10^{4-5}$ $\rm M_{\odot} \, pc^{-3}$ compared to $\sim 1-10$ $\rm M_{\odot} \, pc^{-3}$ in the Solar neighbourhood. This high stellar density favours dynamical encounters between stars that can lead to exchange encounters where neutron stars become part of new binary systems. The exchanges of companions increase the chance of the neutron star going through an accretion phase boosting the number of low mass X-ray binaries (LMXB) that evole into MSPs. Once a neutron star has been recycled into an MSP, it is likely to remain active for a time even longer than the Hubble time due to the low magnetic field.  

Due to the peculiar environment within globular clusters, the possibility of going through multiple stages of recycling and the lack of recent star formation, the majority of pulsars have very fast rotational periods usually smaller than 20 ms. The fastest pulsar known, PSR J1748$-$2446ad\cite{Hessels2006}, with a rotational period of 1.395 ms resides in the globular cluster Terzan 5. Not all of the pulsars in globular clusters are, however, very fast. There is a small population of pulsars where the accretion process was interrupted before completion \cite{Verbunt2014}, possibly by external encounters, with period between 20-100 ms called ‘mildly recycled'.   
Furthermore, there is, in some specific clusters, a small number of slow pulsars with periods over 100 ms with the record of 1004.04 ms being held by B1718$-$19A in NGC 6342 \cite{Lyne1993}. The presence of these apparently ‘young' pulsars is in contradiction with the known formation scenarios and the lack of new star formation. Apart from dynamic disruption of LMXB \cite{Verbunt2014} other possible explanations have been put forward like accretion-induced collapse, direct collisions with a main sequence star \cite{Lyne1996}, or electron capture supernova of an OMgNe white dwarf \cite{Boyles2011}.

The MSPs populations of globular clusters is also quite different from the one in the Galactic disk. 
The percentage of isolated to binary MSPs can vary a lot between clusters with some like NGC 6624 and NGC 6517 having mostly isolated pulsars and others like M62 and M5 having mostly binary pulsars. This is linked to the dynamical parameters of the clusters\cite{Verbunt2014}. 

The multiple dynamical interactions create a population of binary MSPs that are in some cases very different from the Galactic disk population. Eclipsing systems with non-degenerate companions like ‘black widows' and ‘redbacks' are common \cite{Lyne1993,Camilo2000,Ransom2005}. Binary systems can acquire high eccentricities and massive companions through successive interactions or through exchange encounters like M15C\cite{Anderson1990}, NGC 1851A \cite{Freire2004}, NGC 6544B \cite{Lynch2012}, and NGC6652A\cite{Decesar2015} . A very peculiar triple system with a white dwarf and a Jupiter-mass companion has also been observed in a globular cluster \cite{Thorsett1999,Sigurdsson2003}.

\subsection{Science with pulsars in globular clusters}

The scientific achievements made possible by this unique population of pulsars together with the possibility of observing a large number of pulsars in the same telescope beam have made globular clusters a high priority target in the past decades.

The internal composition of neutron stars is determined by the equation-of-state that can be probed by measuring their masses. This can be achieved through pulsar timing by measuring the so-called post-Keplerian parameters, which quantify relativistic orbital effects \cite{Antoniadis2013}. One of these parameters is the relativistic precession of the periastron that is relatively easily accessible in highly eccentric systems, and can be used to place constraints on the total mass of the systeem assuming general relativity. Thanks to this method, some neutron stars with the potential of breaking the current mass limit have been identified\cite{Freire2008,Freire2008b}. The importance of determining neutron star masses is not limited to the value of the maximum mass, also the minimum mass and the distribution of masses can be used to test various formation and recycling scenarios\cite{Ridolfi2019}.

Another avenue of research that can be pursued in globular clusters regards the maximum rotation speed. Currently the record of 1.395 ms is held by J1748-2446ad in Terzan 5\cite{Hessels2006} but even faster pulsars might be present in clusters and set more constraints on the equation of state.

Studies of eclipsing pulsars like ‘black widows' and ‘redbacks' are essential to study the possible evolution scenarios of MSPs and to constrain the properties of the companions\cite{vanKerkwijk2000,Bogdanov2005}.

If we change focus from specific pulsars to the entire population of pulsars in a single cluster, the science perspective broadens to include the properties of the globular clusters themselves. The gravitational effects of the globular clusters on the pulsars are accessible through the derivatives of the rotational period\cite{Phinney1993}. 
With a significant number of pulsars, this can give us access to the structural parameters of the clusters and explore the possibility of the presence of an intermediate mass black hole in the center\cite{Prager2017,Perera2017,Abbate2018}.

Using the dispersion measures (DM) and estimating the 3-d position of the pulsars inside the cluster, it becomes possible to look for diffuse gas. This method has led to the first detection of ionized gas in a globular cluster\cite{Freire2001,Abbate2018}. If we add the information coming from the rotation measure, it becomes possible to probe also the magnetic fields. This led to the possible discovery of a magnetized outflow from the Galaxy in the direction of the globular cluster 47 Tucanae\cite{Abbate2020}.

The numerous scientific possibilities suggest that globular clusters pulsars should still be considered extremely profitable targets for fututre research. Simulations show that a significant population of potentially interesting pulsars is still undiscovered in the clusters\cite{Bagchi2011,Turk2013}. These could include MSP-black hole systems, which would be invaluable for tests of gravitational theories\cite{Liu2014}. However, despite the efforts, the number of new discoveries in the last ten years has been very low. The ones that were discovered were found in archival data using new search techniques \cite{Pan2016,Cadelano2018,Andersen2018}. This suggests that we had reached the sensitivity limit of the current observing facilities. To uncover the numerous and potentially interesting pulsars new and more sensitive facilities are needed. One such facility is FAST\cite{Nan2011} in China that became operative in 2016 and has already discovered 32 pulsars in globular clusters\cite{Pan2020,Wang2020,Pan2021}. Another new facility that recently came online and has the potential to revolutionize the field is MeerKAT\cite{Booth2012} in South Africa.

\section{Advantages of the MeerKAT Radio Telescope}\label{sec:meerkat}

The MeerKAT radio telescope is an interferometer composed of 64 antennas each with a diameter of 13.5 m. The maximum baseline within antennas is 8 km. Thanks to its location in South Africa at a latitude of $-30^{\circ}$, it can observe the entire Southern sky. When observing with all antennas it can reach telescope gains four times higher than Parkes, the only other radio telescope capable of observing pulsars in cm-wavelengths below a declination of $-45^{\circ}$. The gain is also 1.4 times higher than the Green Bank Telescope giving MeerKAT a significant advantage when looking at the central regions of the Galaxy that are observable at high elevations for long hours. 

MeerKAT is currently able to observe in  UHF-band (544-1088 MHz), L-band (856-1712 MHz) showing remarkably RFI-clear bands. In near future also the S-band (1750-3500 MHz) receivers will be available for observations. A description of the capabilities and potentialities of MeerKAT as a pulsar observatory can be found in Bailes et al. 2020\cite{Bailes2020}. 

Of the MeerKAT Large Science Projects (LSPs) that are currently using MeerKAT for observations, two include pulsars as the main target: MeerTIME\cite{Bailes2016} and TRAPUM\cite{Stappers2016}. 

\subsection{MeerTIME}

MeerTIME\footnote{http://www.meertime.org} is an LSP based on timing already known pulsars with the high sensitivity of MeerKAT that started collecting data in early 2019. It has four main scientific goals: timing relativistic binary pulsars in order to measure the masses of the systems and test theories of gravity\cite{Kramer2021}; the Thousand Pulsars Array that monitors a wide variety of pulsars to study their geometries, magnetospheres and interstellar effects \cite{Johnston2020}; a pulsar timing array that monitors high precision pulsars to contribute to the search for nanohertz gravitational waves [Spiewak et al. submitted]; time pulsars in globular clusters\cite{Ridolfi2021}. 

The project plans to observe globular cluster pulsars in order to study the eclipses, measure the masses of the pulsars, and study the properties of the globular clusters themselves. 

MeerTIME can observe up to four tied array beams using the Pulsar Timing User Supplied Equipment (PTUSE) machines as the main data acquisition system. These beams can either be in “timing” mode where a pulsar is folded in real time using an ephemeris or in “search” mode where the data is saved in a “filterbank” style file and can be folded later or searched for new pulsars or single pulses from previously known ones. For both modes the observations are done with very high time resolution (9 $\mu$s), full polarimetric information, and real-time coherent de-dispersion.

Observing globular clusters has the advantage that a large number of pulsars fall within the same beam of the telescope and we are therefore capable of observing them simultaneously if the data is recorded in “search” mode. However, the beam of MeerKAT using the full array with a baseline of 8 km would be too small to observe all of the pulsars in some globular clusters so in most cases only the central 44 antennas with a maximum baseline of 1 km are used.

\subsection{TRAPUM}

The TRAnsients and PUlsars with MeerKAT (TRAPUM\footnote{http://www.trapum.org}) LSP focuses on searching for new pulsars. It searches for pulsars in unidentified Fermi sources, supernova remnants, pulsar wind nebulae, in the Galactic plane, in nearby galaxies and in globular clusters. The globular cluster working group is focused on searching both in clusters with known pulsars (which allows de-dispersion at the known DM of the cluster) and in clusters with no known pulsars. The project started taking data in April 2020.

TRAPUM uses Filterbanking Beam-former User Supplied Equipment (FBFUSE) and the Accelerated Pulsar Search User Supplied Equipment (APSUSE), a 60 node computing cluster, to synthesize up to $\sim 400$ beams and record incoherently de-dispersed search-mode data for each one. This allows us to cover a large portion of the sky (up to $\sim 1 - 4$ arcmin in radius depending on the configuration) while keeping the full sensitivity of MeerKAT. In this way it is possible to cover the entire cluster with one pointing and search for pulsars in the center and in the outskirts of the cluster simultaneously. However, due to the very high data rate, it is not possible to retain full temporal and spectral resolution nor full polarimetric information.

Because of the complementarity of science goals, the globular cluster working groups of MeerTIME and TRAPUM have decided to collaborate sharing observing time and resources. This collaboration has already resulted in two publications \cite{Abbate2020b,Ridolfi2021}.

\section{New discoveries}\label{sec:discoveries}

The joint collaboration has discovered 36 new pulsars in globular clusters\footnote{ the full up-to-date list of discoveries can be found at http://www.trapum.org/discoveries.html}. The discoveries are summarized in Table \ref{table1} and \ref{table2} and can be split chronologically into two groups: the ones made with the PTUSE machine before TRAPUM became operative and the ones made possible by the APSUSE machine of TRAPUM.
 
\begin{table}
\tbl{Properties of the pulsars discovered by MeerTIME/TRAPUM using the PTUSE machine}
{\begin{tabular}{@{}cccc@{}}
\toprule
Pulsar name & Period & DM & notes \\
&  (ms) & (pc cm$^{-3}$) \\
\colrule
Ter 5 an  & 4.8023 & 237.7 &  	Binary \\
M62G	  & 4.6081 & 113.6 &  	Binary \\
NGC 6522D & 5.5369 & 192.7 &  	Isolated \\
NGC 6624G & 6.0912 & 86.20 &  Binary \\
47 Tuc ac & 2.7456 & 24.46 &  Ecl. Binary \\
47 Tuc ad & 3.7460 & 24.41 &  Ecl. Binary \\
NGC 6752F & 8.4854 & 33.20 &  Isolated \\
NGC 6624H & 5.1301 & 86.85 &  Isolated \\
NGC 6440G & 5.2157 & 219.7 &  	Isolated \\
NGC 6440H & 2.8486 & 222.6 &  	Binary \\ 
M28M	  & 9.5689 & 119.23 &	Binary \\
\botrule
\end{tabular}
}
\label{table1}
\end{table}

\begin{table}
\tbl{Properties of the pulsars discovered by MeerTIME/TRAPUM using the APSUSE machine}
{\begin{tabular}{@{}cccc@{}}
\toprule
Pulsar name & Period & DM  & notes \\
&  (ms) & (pc cm$^{-3}$) \\
\colrule
NGC 6752G &	4.7902	 & 33.27	 & Isolated \\    
M28N	  & 3.3429	 & 119.33 & 	Binary \\
NGC 6342B &	2.5683	 & 71.44	 & Isolated \\
NGC 1851B &	2.8162	 & 52.07	 & Isolated \\
NGC 1851C &	5.5648	 & 52.05	 & Isolated \\
NGC 1851D &	4.5543	 & 52.17	 & Binary\\
NGC 1851E &	5.5952	 & 51.95	 & Binary \\
NGC 1851F &	4.3294	 & 51.63	 & Binary \\
NGC 1851G &	3.8028	 & 51.01	 & Binary \\
NGC 1851H &	5.5061	 & 52.26	 & Binary \\
NGC 6752H &	2.0155	 & 33.25	 & Isolated\\
NGC 6752I &	2.6582	 & 33.34	 & Isolated\\
NGC 6624I &	4.3195	 & 87.35	 & Isolated\\
NGC 6624J &	20.8995	 & 86.85	 & Isolated\\
NGC 6441E &	251.1587 & 221.16 & 	Isolated\\
NGC 6441F &	6.0006	 & 228.22 & 	Binary\\
NGC 6441G &	5.3522	 & 229.20 & 	Isolated\\
NGC 1851I &	32.6538	 & 52.42	 & Binary\\
NGC 1851J &	6.6329	 & 52.06	 & Isolated\\
NGC 1851K &	4.6920	 & 51.93	 & Isolated\\
NGC 1851L &	2.9586	 & 51.23	 & Binary\\
NGC 1851M &	4.7977	 & 51.66	 & Isolated\\
NGC 1851N &	5.5679	 & 51.11	 & Isolated\\
J1823-3022 & 2497.72  & 96.9      & Isolated\\
NGC 6624K & 2.7686  & 87.46      & Isolated\\
\botrule
\end{tabular}
}
\label{table2}
\end{table}


\subsection{MeerTIME discoveries}
\begin{figure}[h]
\begin{center}
\includegraphics[width=5in]{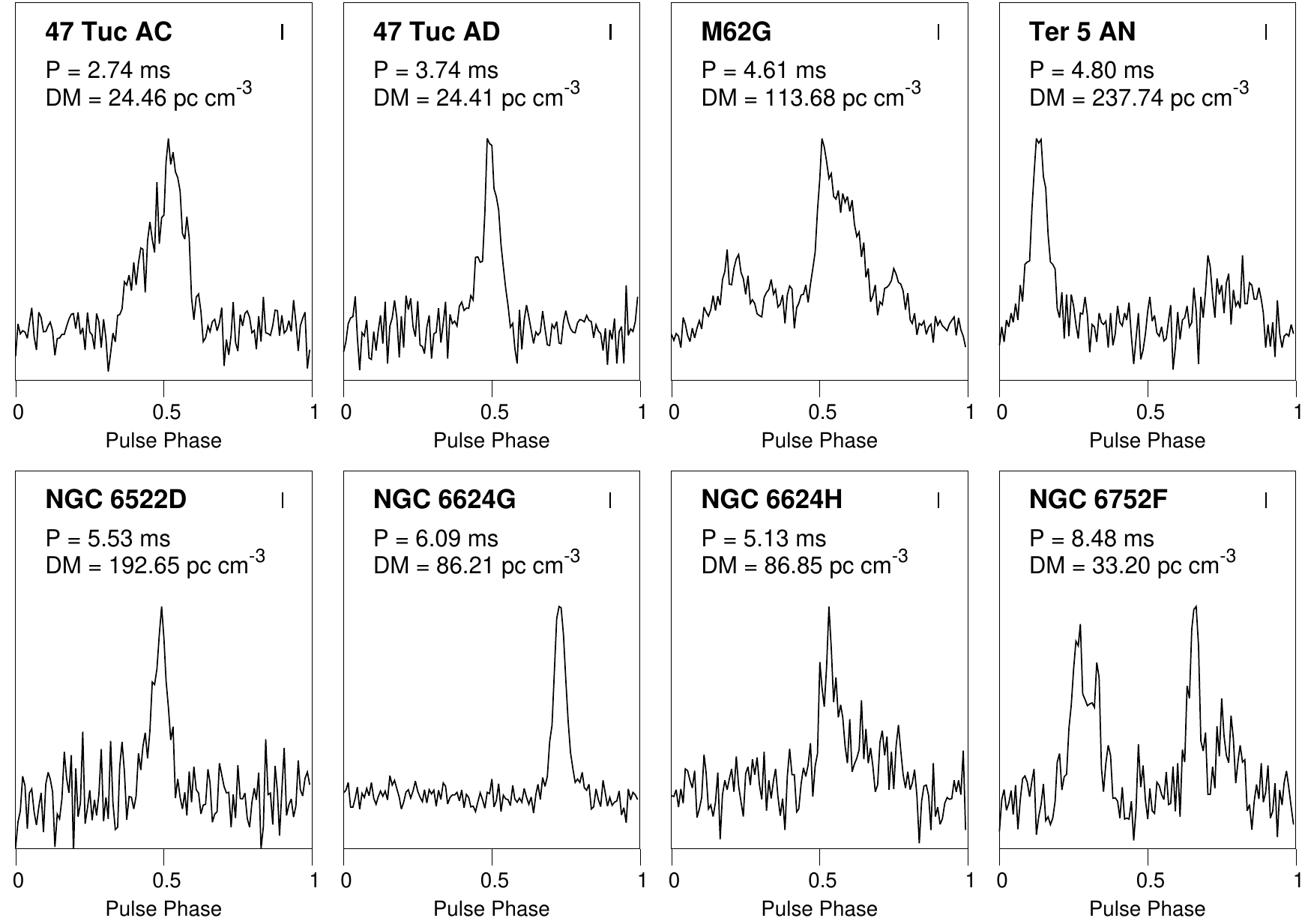}
\end{center}
\caption{Profiles of the eight new pulsar discoveries published in Ridolfi et al. 2021\cite{Ridolfi2021}.}
\label{figure1}
\end{figure}

The first batch of discoveries made with observations taken by MeerTIME is already published\cite{Ridolfi2021}. This publication contains the description and the properties of 8 new millisecond pulsars found in six different globular clusters. Three of them are isolated pulsars while the rest are found in binary systems. The profiles of these pulsars are shown in Figure \ref{figure1}.

These discoveries include two new eclipsing millisecond pulsars 47 Tuc ac and 47 Tuc ad. Fitting for the orbits of these pulsars suggest that 47 Tuc ac is a ‘black widow' pulsar while 47 Tuc ad is a ‘red-back'.  Both of these discoveries were then re detected in previous Parkes observations but due to scintillation effects and eclipses the detections were too few to allow for deriving phase-connected timing solutions.

\begin{figure}[h]
\begin{center}
\includegraphics[width=5in]{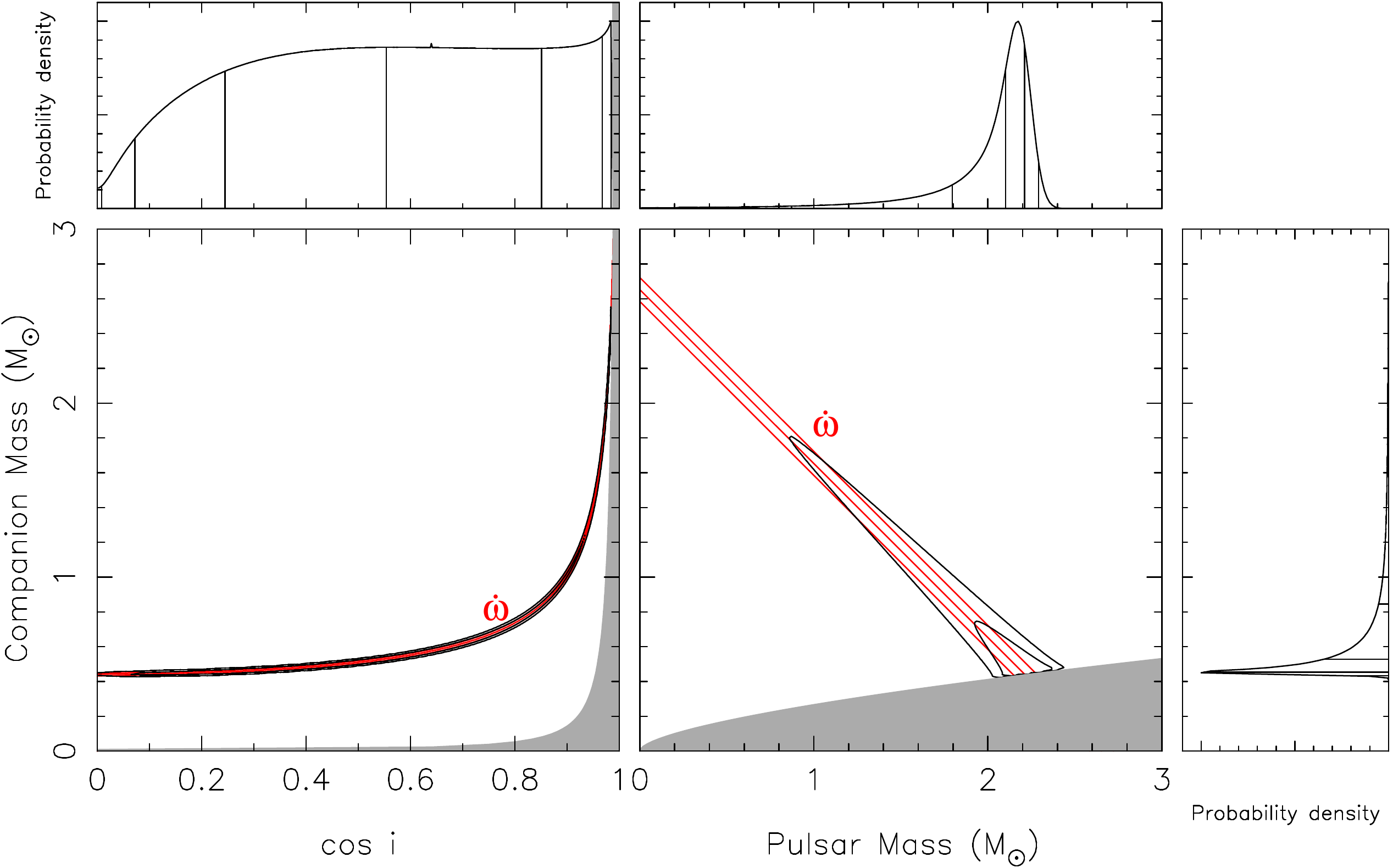}
\end{center}
\caption{Mass-inclination and mass-mass diagrams for NGC 6624G. The main square panels depict the $\cos i$-$M_{\rm c}$ and $M_{\rm c}$-$M_{\rm p}$ planes. The grey areas in the former are excluded by the requirement that the mass of the pulsar is positive, the grey areas in the latter are excluded by the mass function and the requirement that $\sin i \leq  1$. The red lines depict the masses consistent with the measurement of $\dot{\omega}$ and its $\pm  1$-$\sigma$  uncertainties, under the assumption that this effect is dominated by the GR contribution and that GR is the correct theory of gravity.  The contours include 68.3 and 95.4\% of a 2-D probability distribution function (pdf) derived from the $\chi^2$ of \texttt{TEMPO}\ fits that assumed all GR effects to be according to the masses and orbital inclination at each point. The side panels show the probability density functions for the $\cos i$ (top left), $M_{\rm p}$ (top right) and $M_{\rm c}$ (right) derived by marginalizing the aforementioned 2-D pdfs. we obtain a median for the pulsar mass of $2.1$~$\rm M_{\odot}$, but with a tail of probability  that extends to lower masses: there is a 31\% probability of $M_{\rm p} < 2$~$\rm M_{\odot}$\ and a 3.8\% probability of $M_{\rm p} < 1.4$~$\rm M_{\odot}$. Thus, either the pulsar is very massive, or it has a massive companion; the system was likely formed in an exchange encounter. Credits: Ridolfi et al. 2021\cite{Ridolfi2021}.}
\label{figure2}
\end{figure}

Possibly the most interesting pulsar in this set of discoveries is NGC 6624G. This pulsar is found in a binary with orbital period of 1.54 days and eccentricity of 0.38. Because of the very high eccentricity it was possible to use only 11 months of data to measure the rate of advance of periastron $\dot \omega =0.217 \pm 0.004$ deg yr$^{-1}$. Assuming that this advance is caused entirely by relativistic effects and using formulas derived in General Relativity (GR) it is possible to derive the total mass of the system, $M_{tot}= 2.65 \pm 0.07 \rm M_{\odot}$. This parameter alone is not enough to measure precisely the mass of the pulsar itself, see Figure \ref{figure2}, but suggests that the neutron star is particularly heavy. Measurements of further post-keplerian parameters will be needed to accurately determine the mass. The companion mass is also particularly heavy suggesting that it was not the original companion that recycled the pulsar but a new companion gained in an exchange encounter. This is the type of secondary exchange products that we have been looking for.

\begin{figure}[h]
\begin{center}
\includegraphics[width=5in]{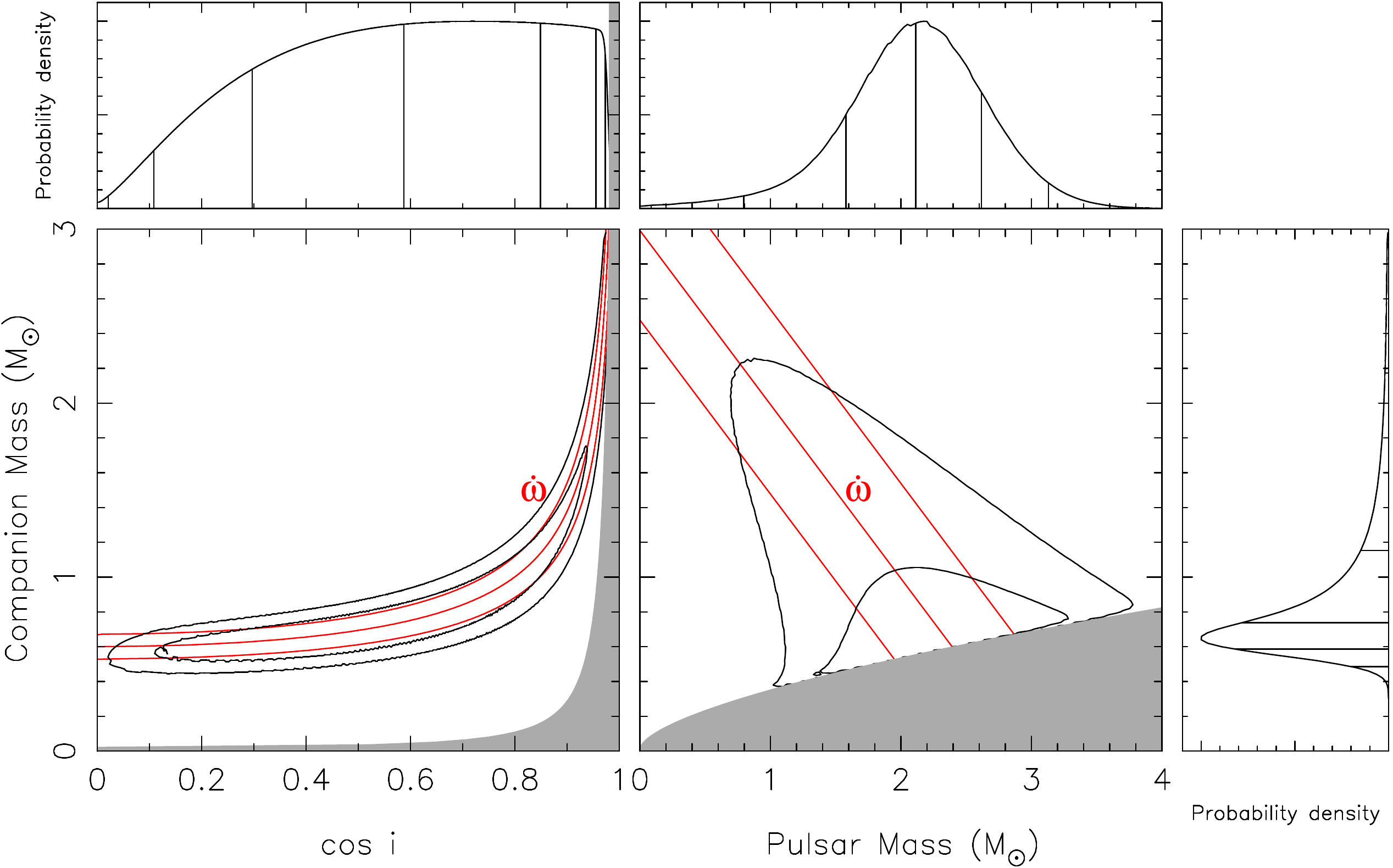}
\end{center}
\caption{Same mass-inclination and mass-mass diagrams as in Figure \ref{figure2} but for Ter 5 an. The estimated median pulsar mass is $2.13$~$\rm M_{\odot}$\ and median companion mass is $0.75$~$\rm M_{\odot}$. Credits: Ridolfi et al. 2021\cite{Ridolfi2021}.}
\label{figure3}
\end{figure}

Another newly discovered pulsar that has the potential of being particularly massive is Ter 5 an. This binary has an eccentricity of 0.0066, two orders of magnitude higher than the average millisecond pulsar binaries found in the Galactic field. It is likely caused by distant stellar encounters in the dense cluster environment. Also here, owing to the recovery of the system in data from the Green Banks Telescope and the resulting long baseline, it was possible to measure an advance of the periastron with a rate of $\dot \omega =0.009 \pm 0.001 $ deg yr$^{-1}$. With the assumption that it fully caused by GR effects, the total mass becomes $M_{tot}= 2.95 \pm 0.52 \rm M_{\odot}$. The mass-inclination and mass-mass diagrams are shown in Figure \ref{figure3}. The uncertainties in this case are higher allowing for a greater range in pulsar mass.

The other pulsars discovered by MeerTIME are in the process of being published. Of these, one is particularly interesting: NGC 6440H. It is a binary pulsar with a very light companion. Assuming a pulsar mass of 1.4 $\rm M_{\odot}$, the companion has a minimum mass of 0.0062 $\rm M_{\odot}$. This very low mass of the companion suggests that it could be brown dwarf or a planet. If the planetary nature of the companion is confirmed, this would be only the second pulsar with such a companion \cite{Thorsett1999,Sigurdsson2003}.

\subsection{TRAPUM discoveries}

Only a small fraction of the observations of globular clusters with the APSUSE machine with TRAPUM have currently been processed. The large volumes of data produced in these observations require a large processing power to be properly analyzed. Despite this, the new pulsar discoveries are already more than double than those made using the PTUSE machines and only the central 44 antennas. The list is shown in Table \ref{table2}.

More than half of these discoveries come from just one globular cluster NGC 1851. This is a core-collapsed cluster that was known to contain only one pulsar NGC 1851A. TRAPUM observations in just the central beam have confirmed the presence of 13 new pulsars. This large population makes NGC 1851 the third globular cluster for number of known pulsars together with M28. Once phase connected ephemerides are determined for all these pulsars it will be possible to probe the potential well and test the presence of gas and magnetic fields within the cluster.

Two of these new discoveries, NGC 1851D and NGC 1851E, are binary systems with very high eccentricities. NGC 1851E has an orbital period of around 7 days and an eccentricity of $e=0.71$. This high value means that in the near future it will be possible to measure the rate of precession of the periastron to high accuracy and therefore the total mass of the system. Early fits to the orbit suggest that, assuming a pulsar mass of 1.4 $\rm M_{\odot}$, the companion has a minimum mass of 1.5 $\rm M_{\odot}$. This suggests that there is a high probability of this being another double neutron star system.

The vast majority of the discoveries are fast spinning MSPs but there are some exceptions: NGC 6624J and NGC 1851I have periods respectively of 20.89 and 32.65 ms, that suggest that they are mildly recycled pulsars. Another exceptions is NGC 6441E, a slow pulsar with a period of 251.15 ms. Particularly interesting is also J1823-3022 with a period of 2497.72 ms. This potentially record breaking pulsar has been found in the outskirts of the globular cluster NGC 6624 and has a DM that is higher than the rest of the pulsars in the cluster. Because of these reasons it is not clear if it is really a member of the cluster or if it is a Galactic disk pulsar positioned along the same line of sight. All of these slow pulsars are in core-collapsed clusters validating the correlation between spin periods and dynamical state of the cluster suggested previously \cite{Verbunt2014}.

These early results from a fraction of the observations suggest that many new discoveries are possible by analyzing the data that has already been captured. 

\section{Future plans}\label{sec:future}

Early observations have showed that the future of the two projects MeerTIME and TRAPUM can prove revolutionary for the science of globular cluster pulsars. 

MeerTIME will focus on timing the already discovered pulsars in order to measure the masses of the most interesting binaries and characterise the eclipses of the ecplising ‘black-widows' and ‘redbacks'. Furthermore, by improving the timing models of all pulsars it will be possible the study the properties of the clusters like the gas content, magnetic field and the presence of intermediate mass black holes.

TRAPUM, on the other hand, will continue to search for new pulsars broadening the scope in order to include clusters with no previously known pulsars. The searches will involve all of the available bands, going from the UHF band, where the pulsars are brighter but the interstellar medium is more noisy to the L band and the S band where the effects of scattering and the interstellar medium are minimized.

In closing, the last few years have shown an almost unprecedented growth in the number of pulsars in globular clusters going from 150 in 2018 to over 230 in 2021. The largest contributor is the MeerKAT telescope with 36 but a very close second is FAST with 32 new discoveries. These numbers are likely to increase in the following years and with them also the possibility of finding even more exotic systems like a double millisecond pulsar binary or a pulsar-black hole binary.

\bibliographystyle{ws-procs961x669}
\bibliography{main}

\begin{thebibliography}{10}

\bibitem{Hessels2006}
J.~W.~T. {Hessels}, S.~M. {Ransom}, I.~H. {Stairs}, P.~C.~C. {Freire}, V.~M.
  {Kaspi} and F.~{Camilo}, {A Radio Pulsar Spinning at 716 Hz}, {\em Science}
  {\bf 311}, 1901 (March 2006).

\bibitem{Verbunt2014}
F.~{Verbunt} and P.~C.~C. {Freire}, {On the disruption of pulsar and X-ray
  binar ies in globular clusters}, {\em Astronomy and Astrophysics} {\bf 561},
  p. A11 (January 2014).

\bibitem{Lyne1993}
A.~G. {Lyne}, J.~D. {Biggs}, P.~A. {Harrison} and M.~{Bailes}, {A long-period
  globular-cluster pulsar in an eclipsing binary system}, {\em Nature} {\bf
  361}, 47 (January 1993).

\bibitem{Lyne1996}
A.~G. {Lyne}, R.~N. {Manchester} and N.~{D'Amico}, {PSR B1745-20 and Young
  Pulsars in Globular Clusters}, {\em Astrophysical Journal Letters} {\bf 460},
  p. L41 (March 1996).

\bibitem{Boyles2011}
J.~{Boyles}, D.~R. {Lorimer}, P.~J. {Turk}, R.~{Mnatsakanov}, R.~S. {Lynch},
  S.~M. {Ransom}, P.~C. {Freire} and K.~{Belczynski}, {Young Radio Pulsars in
  Galactic Globular Clusters}, {\em The Astrophysical Journal} {\bf 742}, p.~51
  (November 2011).

\bibitem{Camilo2000}
F.~{Camilo}, D.~R. {Lorimer}, P.~{Freire}, A.~G. {Lyne} and R.~N. {Manchester},
  {Observations of 20 Millisecond Pulsars in 47 Tucanae at 20 Centimeters},
  {\em The Astrophysical Journal} {\bf 535}, 975 (June 2000).

\bibitem{Ransom2005}
S.~M. {Ransom}, J.~W.~T. {Hessels}, I.~H. {Stairs}, P.~C.~C. {Freire},
  F.~{Camilo}, V.~M. {Kaspi} and D.~L. {Kaplan}, {Twenty-One Millisecond
  Pulsars in Terzan 5 Using the Green Bank Telescope}, {\em Science} {\bf 307},
  892 (February 2005).

\bibitem{Anderson1990}
S.~B. {Anderson}, P.~W. {Gorham}, S.~R. {Kulkarni}, T.~A. {Prince} and
  A.~{Wolszczan}, {Discovery of two radio pulsars in the globular cluster M15},
  {\em Nature} {\bf 346}, 42 (July 1990).

\bibitem{Freire2004}
P.~C. {Freire}, Y.~{Gupta}, S.~M. {Ransom} and C.~H. {Ishwara-Chandra}, {Giant
  Metrewave Radio Telescope Discovery of a Millisecond Pulsar in a Very
  Eccentric Binary System}, {\em Astrophysical Journal Letters} {\bf 606}, L53
  (May 2004).

\bibitem{Lynch2012}
R.~S. {Lynch}, P.~C.~C. {Freire}, S.~M. {Ransom} and B.~A. {Jacoby}, {The
  Timing of Nine Globular Cluster Pulsars}, {\em The Astrophysical Journal}
  {\bf 745}, p. 109 (February 2012).

\bibitem{Decesar2015}
M.~E. {DeCesar}, S.~M. {Ransom}, D.~L. {Kaplan}, P.~S. {Ray} and A.~M.
  {Geller}, {A Highly Eccentric 3.9 Millisecond Binary Pulsar in the Globular
  Cluster NGC 6652}, {\em The Astrophysical Journal Letters} {\bf 807}, p. L23
  (July 2015).

\bibitem{Thorsett1999}
S.~E. {Thorsett}, Z.~{Arzoumanian}, F.~{Camilo} and A.~G. {Lyne}, {The Triple
  Pulsar System PSR B1620-26 in M4}, {\em The Astrophysical Journal} {\bf 523},
  763 (October 1999).

\bibitem{Sigurdsson2003}
S.~{Sigurdsson}, H.~B. {Richer}, B.~M. {Hansen}, I.~H. {Stairs} and S.~E.
  {Thorsett}, {A Young White Dwarf Companion to Pulsar B1620-26: Evidence for
  Early Planet Formation}, {\em Science} {\bf 301}, 193 (July 2003).

\bibitem{Antoniadis2013}
J.~{Antoniadis}, P.~C.~C. {Freire}, N.~{Wex}, T.~M. {Tauris}, R.~S. {Lynch},
  M.~H. {van Kerkwijk}, M.~{Kramer}, C.~{Bassa}, V.~S. {Dhillon}, T.~{Driebe},
  J.~W.~T. {Hessels}, V.~M. {Kaspi}, V.~I. {Kondratiev}, N.~{Langer}, T.~R.
  {Marsh}, M.~A. {McLaughlin}, T.~T. {Pennucci}, S.~M. {Ransom}, I.~H.
  {Stairs}, J.~{van Leeuwen}, J.~P.~W. {Verbiest} and D.~G. {Whelan}, {A
  Massive Pulsar in a Compact Relativistic Binary}, {\em Science} {\bf 340}, p.
  448 (April 2013).

\bibitem{Freire2008}
P.~C.~C. {Freire}, S.~M. {Ransom}, S.~{B{\'e}gin}, I.~H. {Stairs}, J.~W.~T.
  {Hessels}, L.~H. {Frey} and F.~{Camilo}, {Eight New Millisecond Pulsars in
  NGC 6440 and NGC 6441}, {\em The Astrophysical Journal} {\bf 675}, 670 (March
  2008).

\bibitem{Freire2008b}
P.~C.~C. {Freire}, A.~{Wolszczan}, M.~{van den Berg} and J.~W.~T. {Hessels}, {A
  Massive Neutron Star in the Globular Cluster M5}, {\em The Astrophysical
  Journal} {\bf 679}, 1433 (June 2008).

\bibitem{Ridolfi2019}
A.~{Ridolfi}, P.~C.~C. {Freire}, Y.~{Gupta} and S.~M. {Ransom}, {Upgraded Giant
  Metrewave Radio Telescope timing of NGC 1851A: a possible millisecond pulsar
  - neutron star system}, {\em Monthly Notices of the Royal Astronomical
  Society} {\bf 490}, 3860 (December 2019).

\bibitem{vanKerkwijk2000}
M.~H. {van Kerkwijk}, V.~M. {Kaspi}, A.~R. {Klemola}, S.~R. {Kulkarni}, A.~G.
  {Lyne} and D.~{Van Buren}, {Optical Observations of the Binary Pulsar System
  PSR B1718-19: Implications for Tidal Circularization}, {\em The Astrophysical
  Journal} {\bf 529}, 428 (January 2000).

\bibitem{Bogdanov2005}
S.~{Bogdanov}, J.~E. {Grindlay} and M.~{van den Berg}, {An X-Ray Variable
  Millisecond Pulsar in the Globular Cluster 47 Tucanae: Closing the Link to
  Low-Mass X-Ray Binaries}, {\em The Astrophysical Journal} {\bf 630}, 1029
  (September 2005).

\bibitem{Phinney1993}
E.~S. {Phinney}, {Pulsars as Probes of Globular Cluster Dynamics} (January
  1993).

\bibitem{Prager2017}
B.~J. {Prager}, S.~M. {Ransom}, P.~C.~C. {Freire}, J.~W.~T. {Hessels}, I.~H.
  {Stairs}, P.~{Arras} and M.~{Cadelano}, {Using Long-term Millisecond Pulsar
  Timing to Obtain Physical Characteristics of the Bulge Globular Cluster
  Terzan 5}, {\em The Astrophysical Journal} {\bf 845}, p. 148 (August 2017).

\bibitem{Perera2017}
B.~B.~P. {Perera}, B.~W. {Stappers}, A.~G. {Lyne}, C.~G. {Bassa}, I.~{Cognard},
  L.~{Guillemot}, M.~{Kramer}, G.~{Theureau} and G.~{Desvignes}, {Evidence for
  an intermediate-mass black hole in the globular cluster NGC 6624}, {\em
  Monthly Notices of the Royal Astronomical Society} {\bf 468}, 2114 (June
  2017).

\bibitem{Abbate2018}
F.~{Abbate}, A.~{Possenti}, A.~{Ridolfi}, P.~C.~C. {Freire}, F.~{Camilo}, R.~N.
  {Manchester} and N.~{D'Amico}, {Internal gas models and central black hole in
  47 Tucanae using millisecond pulsars}, {\em Monthly Notices of the Royal
  Astronomical Society} {\bf 481}, 627 (November 2018).

\bibitem{Freire2001}
P.~C. {Freire}, M.~{Kramer}, A.~G. {Lyne}, F.~{Camilo}, R.~N. {Manchester} and
  N.~{D'Amico}, {Detection of Ionized Gas in the Globular Cluster 47 Tucanae},
  {\em The Astrophysical Journal} {\bf 557}, L105 (August 2001).

\bibitem{Abbate2020}
F.~{Abbate}, A.~{Possenti}, C.~{Tiburzi}, E.~{Barr}, W.~{van Straten},
  A.~{Ridolfi} and P.~{Freire}, {Constraints on the magnetic field in the
  Galactic halo from globular cluster pulsars}, {\em Nature Astronomy} {\bf 4},
  704 (March 2020).

\bibitem{Bagchi2011}
M.~{Bagchi}, D.~R. {Lorimer} and J.~{Chennamangalam}, {Luminosities of recycled
  radio pulsars in globular clusters}, {\em Monthly Notices of the Royal
  Astronomical Society} {\bf 418}, 477 (November 2011).

\bibitem{Turk2013}
P.~J. {Turk} and D.~R. {Lorimer}, {An empirical Bayesian analysis applied to
  the globular cluster pulsar population}, {\em Monthly Notices of the Royal
  Astronomical Society} {\bf 436}, 3720 (December 2013).

\bibitem{Liu2014}
K.~{Liu}, R.~P. {Eatough}, N.~{Wex} and M.~{Kramer}, {Pulsar-black hole
  binaries: prospects for new gravity tests with future radio telescopes}, {\em
  Monthly Notices of the Royal Astronomical Society} {\bf 445}, 3115 (December
  2014).

\bibitem{Pan2016}
Z.~{Pan}, G.~{Hobbs}, D.~{Li}, A.~{Ridolfi}, P.~{Wang} and P.~{Freire},
  {Discovery of two new pulsars in 47 Tucanae (NGC 104)}, {\em Monthly Notices
  of the Royal Astronomical Society} {\bf 459}, L26 (June 2016).

\bibitem{Cadelano2018}
M.~{Cadelano}, S.~M. {Ransom}, P.~C.~C. {Freire}, F.~R. {Ferraro}, J.~W.~T.
  {Hessels}, B.~{Lanzoni}, C.~{Pallanca} and I.~H. {Stairs}, {Discovery of
  Three New Millisecond Pulsars in Terzan 5}, {\em The Astrophysical Journal}
  {\bf 855}, p. 125 (March 2018).

\bibitem{Andersen2018}
B.~C. {Andersen} and S.~M. {Ransom}, {A Fourier Domain
  {\textquotedblleft}Jerk{\textquotedblright} Search for Binary Pulsars}, {\em
  The Astrophysical Journal Letters} {\bf 863}, p. L13 (August 2018).

\bibitem{Nan2011}
R.~{Nan}, D.~{Li}, C.~{Jin}, Q.~{Wang}, L.~{Zhu}, W.~{Zhu}, H.~{Zhang},
  Y.~{Yue} and L.~{Qian}, {The Five-Hundred Aperture Spherical Radio Telescope
  (fast) Project}, {\em International Journal of Modern Physics D} {\bf 20},
  989 (January 2011).

\bibitem{Pan2020}
Z.~{Pan}, S.~M. {Ransom}, D.~R. {Lorimer}, W.~C. {Fiore}, L.~{Qian}, L.~{Wang},
  B.~W. {Stappers}, G.~{Hobbs}, W.~{Zhu}, Y.~{Yue}, P.~{Wang}, J.~{Lu},
  K.~{Liu}, B.~{Peng}, L.~{Zhang} and D.~{Li}, {The FAST Discovery of an
  Eclipsing Binary Millisecond Pulsar in the Globular Cluster M92 (NGC 6341)},
  {\em The Astrophysical Journal Letters} {\bf 892}, p.~L6 (March 2020).

\bibitem{Wang2020}
L.~{Wang}, B.~{Peng}, B.~W. {Stappers}, K.~{Liu}, M.~J. {Keith}, A.~G. {Lyne},
  J.~{Lu}, Y.-Z. {Yu}, F.~{Kou}, J.~{Yan}, P.~{Jiang}, C.~{Jin}, D.~{Li},
  Q.~{Li}, L.~{Qian}, Q.~{Wang}, Y.~{Yue}, H.~{Zhang}, S.~{Zhang}, Y.~{Zhu} and
  {FAST Collaboration}, {Discovery and Timing of Pulsars in the Globular
  Cluster M13 with FAST}, {\em The Astrophysical Journal} {\bf 892}, p.~43
  (March 2020).

\bibitem{Pan2021}
Z.~{Pan}, L.~{Qian}, X.~{Ma}, K.~{Liu}, L.~{Wang}, J.~{Luo}, Z.~{Yan},
  S.~{Ransom}, D.~{Lorimer}, D.~{Li} and P.~{Jiang}, {FAST Globular Cluster
  Pulsar Survey: Twenty-four Pulsars Discovered in 15 Globular Clusters}, {\em
  The Astrophysical Journal Letters} {\bf 915}, p. L28 (July 2021).

\bibitem{Booth2012}
R.~S. {Booth} and J.~L. {Jonas}, {An Overview of the MeerKAT Project}, {\em
  African Skies} {\bf 16}, p. 101 (March 2012).

\bibitem{Bailes2020}
M.~{Bailes}, A.~{Jameson}, F.~{Abbate}, E.~D. {Barr}, N.~D.~R. {Bhat},
  L.~{Bondonneau}, M.~{Burgay}, S.~J. {Buchner}, F.~{Camilo}, D.~J. {Champion},
  I.~{Cognard}, P.~B. {Demorest}, P.~C.~C. {Freire}, T.~{Gautam}, M.~{Geyer},
  J.~M. {Griessmeier}, L.~{Guillemot}, H.~{Hu}, F.~{Jankowski}, S.~{Johnston},
  A.~{Karastergiou}, R.~{Karuppusamy}, D.~{Kaur}, M.~J. {Keith}, M.~{Kramer},
  J.~{van Leeuwen}, M.~E. {Lower}, Y.~{Maan}, M.~A. {McLaughlin}, B.~W.
  {Meyers}, S.~{Os{\l}owski}, L.~S. {Oswald}, A.~{Parthasarathy},
  T.~{Pennucci}, B.~{Posselt}, A.~{Possenti}, S.~M. {Ransom}, D.~J. {Reardon},
  A.~{Ridolfi}, C.~T.~G. {Schollar}, M.~{Serylak}, G.~{Shaifullah},
  M.~{Shamohammadi}, R.~M. {Shannon}, C.~{Sobey}, X.~{Song}, R.~{Spiewak},
  I.~H. {Stairs}, B.~W. {Stappers}, W.~{van Straten}, A.~{Szary},
  G.~{Theureau}, V.~{Venkatraman Krishnan}, P.~{Weltevrede}, N.~{Wex}, T.~D.
  {Abbott}, G.~B. {Adams}, J.~P. {Burger}, R.~R.~G. {Gamatham}, M.~{Gouws},
  D.~M. {Horn}, B.~{Hugo}, A.~F. {Joubert}, J.~R. {Manley}, K.~{McAlpine},
  S.~S. {Passmoor}, A.~{Peens-Hough}, Z.~R. {Ramudzuli}, A.~{Rust}, S.~{Salie},
  L.~C. {Schwardt}, R.~{Siebrits}, G.~{Van Tonder}, V.~{Van Tonder} and M.~G.
  {Welz}, {The MeerKAT telescope as a pulsar facility: System verification and
  early science results from MeerTime}, {\em Publications of the Astronomical
  Society of Australia} {\bf 37}, p. e028 (July 2020).

\bibitem{Bailes2016}
M.~{Bailes}, E.~{Barr}, N.~D.~R. {Bhat}, J.~{Brink}, S.~{Buchner}, M.~{Burgay},
  F.~{Camilo}, D.~{Champion}, J.~{Hessels}, A.~{Jameson}, S.~{Johnston},
  A.~{Karastergiou}, R.~{Karuppusamy}, V.~{Kaspi}, M.~{Keith}, M.~{Kramer},
  M.~{McLaughlin}, K.~{Moodley}, S.~{Oslowski}, A.~{Possenti}, S.~{Ransom},
  F.~{Rasio}, J.~{Sievers}, M.~{Serylak}, B.~{Stappers}, I.~{Stairs},
  G.~{Theureau}, W.~{van Straten}, P.~{Weltevrede} and N.~{Wex}, {MeerTime -
  the MeerKAT Key Science Program on Pulsar Timing} (January 2016).

\bibitem{Stappers2016}
B.~{Stappers} and M.~{Kramer}, {An Update on TRAPUM} (January 2016).

\bibitem{Kramer2021}
M.~{Kramer}, I.~H. {Stairs}, V.~{Venkatraman Krishnan}, P.~C.~C. {Freire},
  F.~{Abbate}, M.~{Bailes}, M.~{Burgay}, S.~{Buchner}, D.~J. {Champion},
  I.~{Cognard}, T.~{Gautam}, M.~{Geyer}, L.~{Guillemot}, H.~{Hu}, G.~{Janssen},
  M.~E. {Lower}, A.~{Parthasarathy}, A.~{Possenti}, S.~{Ransom}, D.~J.
  {Reardon}, A.~{Ridolfi}, M.~{Serylak}, R.~M. {Shannon}, R.~{Spiewak},
  G.~{Theureau}, W.~{van Straten}, N.~{Wex}, L.~S. {Oswald}, B.~{Posselt},
  C.~{Sobey}, E.~D. {Barr}, F.~{Camilo}, B.~{Hugo}, A.~{Jameson},
  S.~{Johnston}, A.~{Karastergiou}, M.~{Keith} and S.~{Os{\l}owski}, {The
  relativistic binary programme on MeerKAT: science objectives and first
  results}, {\em Monthly Notices of the Royal Astronomical Society} {\bf 504},
  2094 (June 2021).

\bibitem{Johnston2020}
S.~{Johnston}, A.~{Karastergiou}, M.~J. {Keith}, X.~{Song}, P.~{Weltevrede},
  F.~{Abbate}, M.~{Bailes}, S.~{Buchner}, F.~{Camilo}, M.~{Geyer}, B.~{Hugo},
  A.~{Jameson}, M.~{Kramer}, A.~{Parthasarathy}, D.~J. {Reardon}, A.~{Ridolfi},
  M.~{Serylak}, R.~M. {Shannon}, R.~{Spiewak}, W.~{van Straten},
  V.~{Venkatraman Krishnan}, F.~{Jankowski}, B.~W. {Meyers}, L.~{Oswald},
  B.~{Posselt}, C.~{Sobey}, A.~{Szary} and J.~{van Leeuwen}, {The
  Thousand-Pulsar-Array programme on MeerKAT - I. Science objectives and first
  results}, {\em Monthly Notices of the Royal Astronomical Society} {\bf 493},
  3608 (April 2020).

\bibitem{Ridolfi2021}
A.~{Ridolfi}, T.~{Gautam}, P.~C.~C. {Freire}, S.~M. {Ransom}, S.~J. {Buchner},
  A.~{Possenti}, V.~{Venkatraman Krishnan}, M.~{Bailes}, M.~{Kramer}, B.~W.
  {Stappers}, F.~{Abbate}, E.~D. {Barr}, M.~{Burgay}, F.~{Camilo},
  A.~{Corongiu}, A.~{Jameson}, P.~V. {Padmanabh}, L.~{Vleeschower}, D.~J.
  {Champion}, W.~{Chen}, M.~{Geyer}, A.~{Karastergiou}, R.~{Karuppusamy},
  A.~{Parthasarathy}, D.~J. {Reardon}, M.~{Serylak}, R.~M. {Shannon} and
  R.~{Spiewak}, {Eight new millisecond pulsars from the first MeerKAT globular
  cluster census}, {\em Monthly Notices of the Royal Astronomical Society} {\bf
  504}, 1407 (June 2021).

\bibitem{Abbate2020b}
F.~{Abbate}, M.~{Bailes}, S.~J. {Buchner}, F.~{Camilo}, P.~C.~C. {Freire},
  M.~{Geyer}, A.~{Jameson}, M.~{Kramer}, A.~{Possenti}, A.~{Ridolfi},
  M.~{Serylak}, R.~{Spiewak}, B.~W. {Stappers} and V.~{Venkatraman Krishnan},
  {Giant pulses from J1823-3021A observed with the MeerKAT telescope}, {\em
  Monthly Notices of the Royal Astronomical Society} {\bf 498}, 875 (October
  2020).

\end{thebibliography}

\end{document}